\documentclass[twocolumn, dvipsnames]{aastex62}

\graphicspath{{./}{figures/}}

\usepackage{amssymb}
\usepackage{amsmath}

\shorttitle{Similarity of gamma-ray pulsar light curves }
\shortauthors{Garc\'ia \& Torres}

\begin{document}

\title{Quantitative exploration of the similarity of gamma-ray pulsar light curves}

\email{crodriguez@ice.csic.es}
\email{dtorres@ice.csic.es}

\author{C. R. Garc\'{i}a}
\affiliation{Institute of Space Sciences (ICE, CSIC), Campus UAB, Carrer de Can Magrans s/n, 08193 Barcelona, Spain}
\affiliation{Institut d’Estudis Espacials de Catalunya (IEEC), 08034 Barcelona, Spain}

\author{Diego F. Torres}
\affiliation{Institució Catalana de Recerca i Estudis Avançats (ICREA), E-08010 Barcelona, Spain }
\affiliation{Institute of Space Sciences (ICE, CSIC), Campus UAB, Carrer de Can Magrans s/n, 08193 Barcelona, Spain}
\affiliation{Institut d’Estudis Espacials de Catalunya (IEEC), 08034 Barcelona, Spain}

\begin{abstract}
We introduce and apply a methodology based on dynamic time warping (DTW) to compare the whole set of gamma-ray light curves reported in the Third {\it Fermi}-Large Area Telescope Pulsar Catalogue. 
Our method allows us to quantitatively measure the degree of global similarity between two light curves beyond comparing indicators such as how many peaks there are,  which is their separation, width, and height. 
Once the morphology of the light curve is showcased via background subtraction, min-max scaler normalization, and rotations are considered to take into account that phase 0 is arbitrary, the level of detail with which light curves of different pulsars appear is revealed. 
In many cases their similarity is striking and occurs disregarding any other timing, physical, or spectral property.
In particular, some MSPs and young pulsars share detailed light curve morphology.
\end{abstract}

\keywords{pulsars: general, stars: neutron, methods: data analysis
}

\section{Introduction} 
\label{sec:intro}

Up to now, comparing light curves of two or more gamma-ray pulsars has mostly been an artisanal business. 
One would usually compare how similar two light curves are to the eye of the beholder, mentally disregarding the different flux levels.
This is, however, not always possible, as the light curve count rates can vary from a few to more than a thousand for the same bin of phase. 
An improvement over this comes by considering global light curve indicators like how many peaks there are, what their separation is, and their width and height. 
This is not as obvious as it seems for several reasons.
First, a solid definition of what is a light curve peak is lacking. 
Any characterization of light curves would inherently depend on this prior definition.
Secondly, light curves with two peaks of different heights can, in fact, have an equal relative height between them, or light curves with a different peak separation can be fully similar otherwise.
It is clear then that the former indicators do not exhaust the need for a similarity assessment and do not provide a quantitative measurement for it.

Recent theoretical studies of light curve morphology, see \cite{Iniguez2024, Cerutti2024}, emphasize the importance of having such similarity measurements.
Results from theoretical models imply that light curve properties are mostly related to geometry and do not correlate with other physical parameters of the pulsars or their magnetospheres.
\cite{Iniguez2024}, in particular, have demonstrated that pulsars with different timing and spectral parameters have similar radiation skymaps. 
This implies the likely similarity among their magnetospheres and that the geometry widely dominates over the spectra in shaping the light curves. 
If so, considering a sample of light curves generated for the same pulsar but for different geometries and observers should relate to exploring the population of all observed pulsars.
Thus, knowing to what extent pulsars with different properties have similar light curves is essential to advance our understanding of pulsar magnetospheres.

Dynamic time warping (DTW) is a well-known method to compare time series and find similar behaviors, even if varying in speed or have other temporal differences.
\cite{Bellman_1959, Bellman_1961} introduced the concept of dynamic programming, the mathematical framework on which DTW is based.
It has been used for speech recognition \citep{Vintsyuk_1968, Itakura_75, Sakoe1978, Myers_1980, Myers_1981, Godin_1989, Rabiner_1993}, human activity classification \citep{Holt_2007, Sempena_2011}, music and motions analysis \citep{Müller_2007}, signature or fingerprint recognition \citep{Munich_1999, Kovacs-Vajna_2000}, shape matching \citep{Bartolini_2005}, bioinformatics \citep{Aach_2001}, chemical engineering \citep{Dai_2011}, or medicine \citep{Kiani_2017, Gogolou_2019}.  
Beyond these applications, it has also been commonly used to analyze economic trends in stock markets and financial time series, particularly for predicting economic behavior \cite{Tasneem_2017, Franses_2020, Arya_2021}.
It is based on finding the best dynamic alignment, also called the optimal warping path \citep{Keogh_2005}.
Details on how it is computed are briefly given next. 
DTW has also recently started to be used in astrophysics. 
Examples are the similarity between light curves of Gamma-ray bursts that offers insight into their central engine activities \cite{Zhang_2016}, the reconstruction of gravitational wave signals from core-collapse supernovae \cite{Suvorova2019}, or the assessment of solar wind time series \cite{Samara2022}. Recently, \cite{Vohl2024} has used an unsupervised method involving DTW to sort radio pulsars by profile-shape similarity using graph topology. 
Here, for the first time, we propose and use a methodology based on DTW to quantitatively assess the degree of similarity between two gamma-ray light curves and apply it to the whole set of light curves in the 3PC.

\section{Computing dynamic time warping}
\label{part_methods}

Consider two time series $X$ and $Y$ of  length $n$ and $m$, 
\begin{equation}  
    \begin{aligned}
        X &= [x_1, x_2, \dots, x_i, \dots, x_n] \\
        Y &= [y_1, y_2, \dots, y_j, \dots, y_m].
    \end{aligned}
    \label{eq: time_series}
\end{equation}
The indices $(i,j)$ run along the elements of each time series, respectively ($i=1..n$, $j=1..m$).
To be considered a warping path, i.e., a set of connected pairs of points $(x_i, y_j)$ covering all the elements of both series, we shall require three conditions that ensure efficient, meaningful, and temporally consistent alignments, see eg., \cite{Berndt_1994}.
First, each of the time series's initial and final elements faces each other, so that any warping path has constrained endpoints, that is, $i_{1}=1$, $j_{1}=1$, and $i_{k}=n$, $j_{k}=m$.
We shall also require that no elements of the series that break the temporal order be connected, i.e., no backward steps. Said mathematically, $i_{k-1} \leq i_{k}$ and $j_{k-1} \leq j_{k}$.
Finally, we shall also request that the elements of the series being connected do so without gaps, or simply that all elements in the series are involved in the path, $i_{k}-i_{k-1} \leq 1$ and $j_{k}-j_{k-1} \leq 1$.
Taking these conditions (referred to as boundary, monotonicity, and continuity) into account, the DTW is computed as (see e.g., \cite{Hailin_2014}),
\begin{eqnarray}
\text{DTW}(X, Y) = 
\min_{\pi\in \mathcal{A}(X,Y)}
\Big(\sum_{(i,j)\in\pi}d(x_{i},y_{j})^{2}\Big)^{\frac{1}{2}}
,\label{eq: DTW_optimized}
\end{eqnarray}
where $\pi$ represents an explored warping path that pertains to the set $\mathcal{A}$ of all paths 
fulfilling the above-referred conditions.
The DTW value from Eq. (\ref{eq: DTW_optimized}) is the cost of the optimal warping path ($\pi_{\mathrm{o}}$), that is, the minimum value obtained after adding the distance $d$ between all pairs $(i,j)$ of $\pi_{\mathrm{o}}$.
Finally, in addition to comparing light curves using DTW, we shall also use a direct computation of the Euclidean distance (ED) when the length in the two-time series is the same or they have been correspondingly rebinned to the same. 
The Appendix provides a worked-out example that will help generate intuition regarding these definitions.
The Appendix also provides details regarding our numerical implementation and its cost.

\section{{\it Fermi}-LAT 3PC Light curves }
\label{part_set_data}

The set of time series herein considered comprises 294 light curves associated with the pulsars reported in the {\it Fermi}-LAT Third Pulsar Catalog (3PC, \cite{Fermi3PC}).
In this sample, there are 143 millisecond pulsars (MSPs) with periods shorter than 10 ms ($P<10$ ms).
We consider the pulsar's weighted counts above $E\geq 100$ MeV, subtracting the background level for each pulsar as reported in the 3PC.
When the resulting value is lower than zero, we set it to zero (no pulsar contribution above the background). 
As we are interested in comparing the morphology of the light curves, they should have the same scale; thus we consider the \textit{Min-Max scaler} after background subtraction, i.e., all light curves are transformed to lie between 0 and 1 via
\begin{eqnarray}
X^{\dag}=\frac{x_i-x_{\text{min}}}{x_{\text{max}}-x_{\text{min}}}.
\label{eq: min-max_scaler}
\end{eqnarray}
Gamma-ray light curves have different sizes, i.e., are discretized in a different number of bins ($N$).
The 3PC contains 48 light curves described with 25, 116 with 50, 98 with 100, 27 with 200, 4 with 400, and 1 with 800 bins.
These bins correspond to the rotational phase of a pulsar ($\phi \in (0,1)$).
However, as $\phi$=0 is arbitrary, we shall consider all possible phase-shifts, getting a DTW for each using Eq. (\ref{eq: DTW_optimized}).
The minimum value of DTW will measure how similar two light curves are once the most favorable phase shift between them is chosen.
%

\section{Results} 
\label{results}

\subsection{Comparing light curves with the same number of bins via ED}
\label{ed_part_dtw}

\begin{figure*}
  \includegraphics[width=1\textwidth]{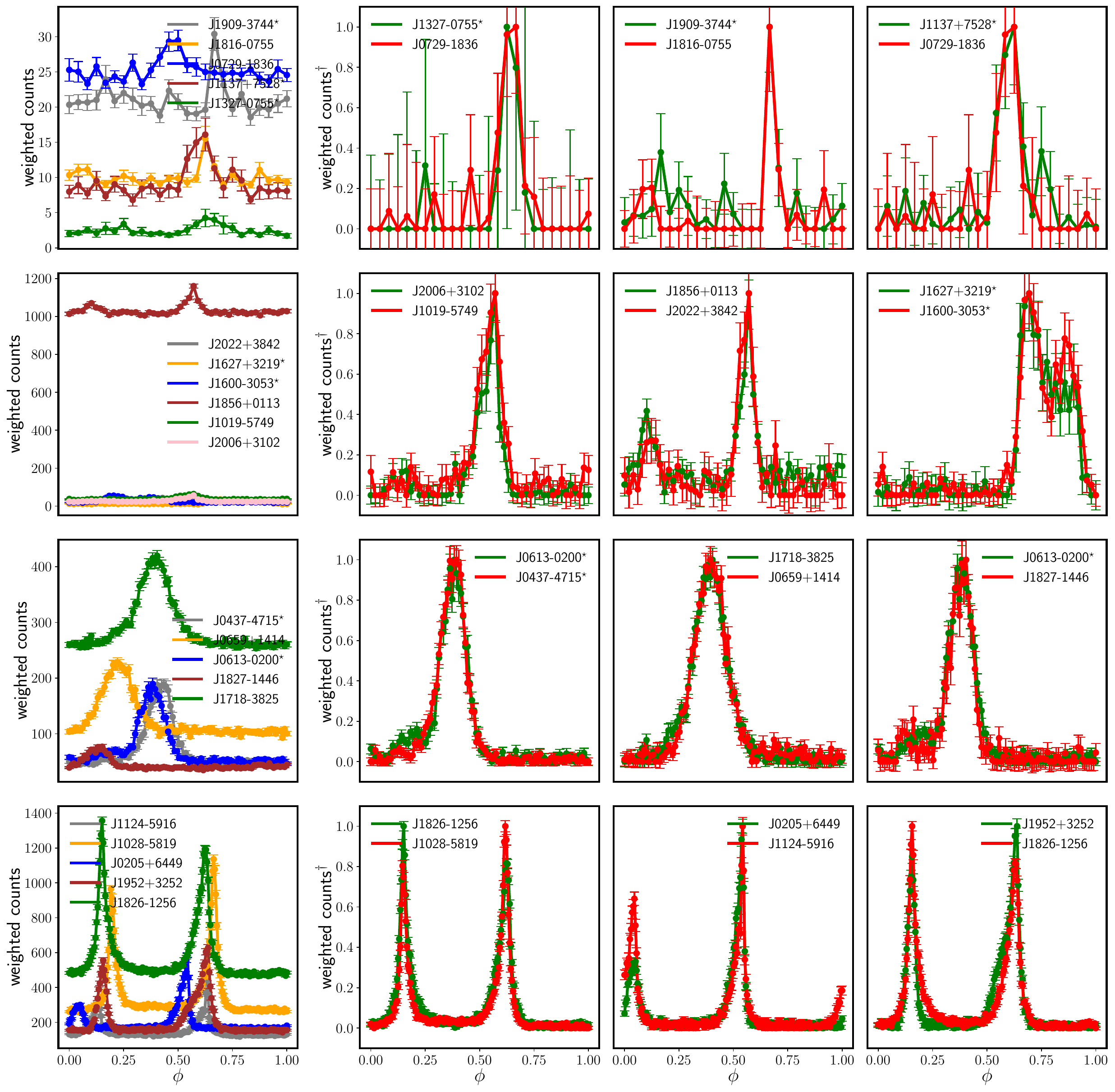}
  \centering
\caption{Each row shows the 3 most similar pairs of light curves according to $\mathrm{ED}$ for light curves having 25, 50, 100, and 200 bins.
The plot in the first column of each row shows the 3PC data directly before processing.
MSPs are denoted with a star on the pulsar name. }

       \label{fig: ranking_ED}
\end{figure*}

Figure \ref{fig: ranking_ED} shows the three most similar pairs of light curves as quantified by ED. 
Figure \ref{fig: ranking_ED} also shows how these light curves look as read from the 3PC before processing, i.e., before background subtraction, before the Eq. \ref{eq: min-max_scaler} is used to compare the shape disregarding different flux levels, and before rotations are performed to produce the best alignment leading to the minimum value of ED. 
The morphological similarity of these pairs of light curves is striking. 
In particular, MSPs and young pulsars share detailed light curve morphology.
The pulsar spin period ($P$) and the spin period derivative ($\dot{P}$), and the quantities derived from them using the rotating dipole model, such as the surface magnetic flux density  ($B_{s}$), the magnetic field at the light cylinder ($B_{lc}$), and the spin-down energy loss rate ($\dot{E}_{sd}$), can be very different from one pulsar to another, even when their light curves are very similar. 
Differences up a factor of 70 in $\dot{E}_{sd}$, of 4 in $B_{s}$, of 30 in $B_{lc}$ can be found even between young pulsars pairs sharing light curve morphology. 
The differences in parameters, when the pair mixes MSPs and young pulsars, can be much larger, e.g., reaching up to a factor more than $3\times 10^4$ in
$B_{s}$.
The gamma-ray emission properties, as reported in the 3PC, e.g., phase-averaged integral energy flux in the 0.1 to 300 GeV energy band, $G_{100}$ ($F_{E>100\mathrm{MeV}}$), and the energy of the maximum peak of the SED ($E_p$) also span a large range, representing very different gamma-ray spectra.
Light curve similarity is immune to such differences as well.

\subsection{Comparing light curves of the same size via DTW}

We also use the DTW methodology to analyze whether it brings something new compared to ED.
Results for the set of most similar, equal-size light curves under DTW are shown in Fig. \ref{fig: ranking_dtwbyed}.
We recall that the cumulative cost matrix in DTW ($D$, see details in the Appendix, especially Eq. (\ref{eq: MatrixDcum2})) includes the ED comparison.
Thus, all paths considered via ED are also considered when using DTW. 
In some cases, the DTW ranking of the three most similar pairs for each subset of equal-$N$ light curves is the same as those in ED. In general, this is not the case. 
The Appendix example also comments on how this happens.

The similarity of the pulsar light curves showcased by DTW goes beyond simple visual inspection (which is what ED matches more closely).
It does not focus on e.g., having similar peaks separated by the same phase interval, which our eye would do directly if we could rotate all light curves to consider all possible shifts. Instead, it focuses on the global morphology. 
Consider, for example, the case of the group of $N=200$ bins. 
The DTW ranking chooses the most similar pair, J0614-3329, and J2032+4127, which is not even quoted in the first three pairs ranked by ED. 
The pulsar pairs selected by ED are visually appealing to the eye because the $\phi$ separation is the same despite the height of each peak is different. Under DTW, however, these do not rank as highly, as this method prioritizes overall morphological similarity rather than strict phase alignment.
One example of this is the top pair ranked by DTW, which has a slightly different phase separation between peaks (which for DTW is not a problem but would increase ED significantly) and an almost exact match of peak heights. DTW places more emphasis on peak values, when comparing these light curves, whereas ED places more emphasis on phasing.
As before, they involve pulsars of all kinds with significantly different physical and spectral parameters.

\begin{figure*}
  \includegraphics[width=\textwidth]{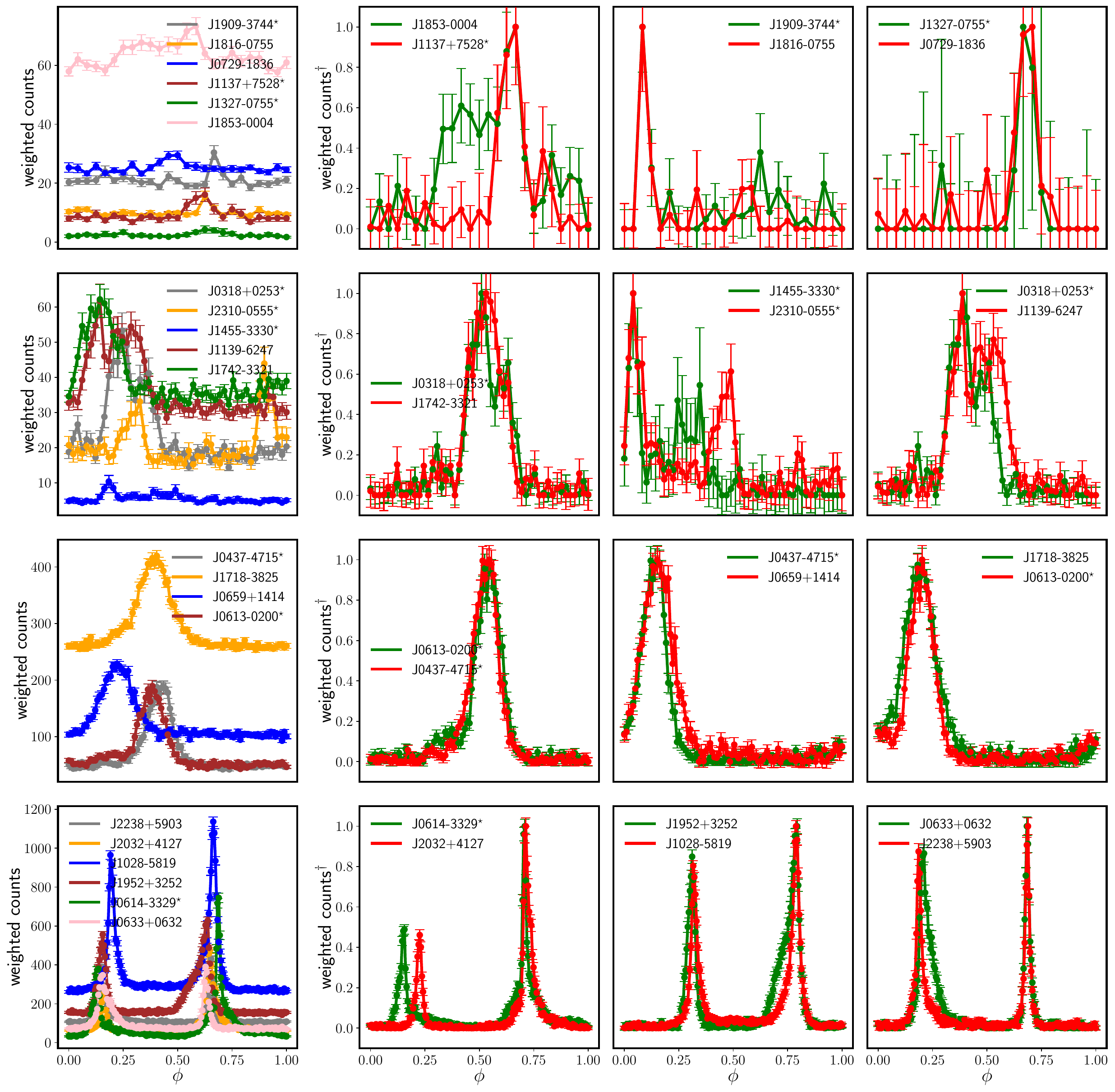}
  \centering
\caption{The 3 most similar pairs of light curves according to $\mathrm{DTW}$ for light curves having 25, 50, 100, and 200 bins.
The plot in the first column of each row shows the 3PC data directly, before processing.
MSPs are denoted with a star on the pulsar name. }
       \label{fig: ranking_dtwbyed}
\end{figure*}

\subsection{Caveats of ED and rebinning the whole sample}
\label{caveats_ED}
ED has a restricted application to light curves of equal size. 
This restriction would leave aside the best-described light curves from most similarity comparisons unless rebinned to a lower resolution, losing details. 
Rebinning, thus, (unless rebinning to the worst-resolved light curve, with $N=25$) will produce the loss of significant portions of the sample.
DTW remains stable despite rebinning.
For instance, over the sample of light curves rebinned to 100 bins (formed by 130 light curves, as those with $N<100$ bins are ignored), DTW rankings of the 5 most similar light curves from any given one retain 3 pulsars ranked also when no rebinning is applied in 85\% of the cases.
If we look at the two best matches, at least one is reproduced by both situations (with and without rebinning) in 88\% of the cases.
In addition of reducing the sample, we have shown that ED is bound to produce large measures (signaling dissimilarity) even in cases where the overall structure is the same.
In particular, this will happen when two or more peaks are not precisely aligned, there is a slight difference in phase separation, or the peak widths are slightly different.
When rebinning, as the resolution at which the comparison will be made will, at most, be that of the lesser-sampled light curve, these issues will appear more often.
DTW solves both of these problems at once.

\subsection{Using DTW to compare all light curves}
\label{part_dtw}

\begin{figure}
  \includegraphics[width=0.4\textwidth]{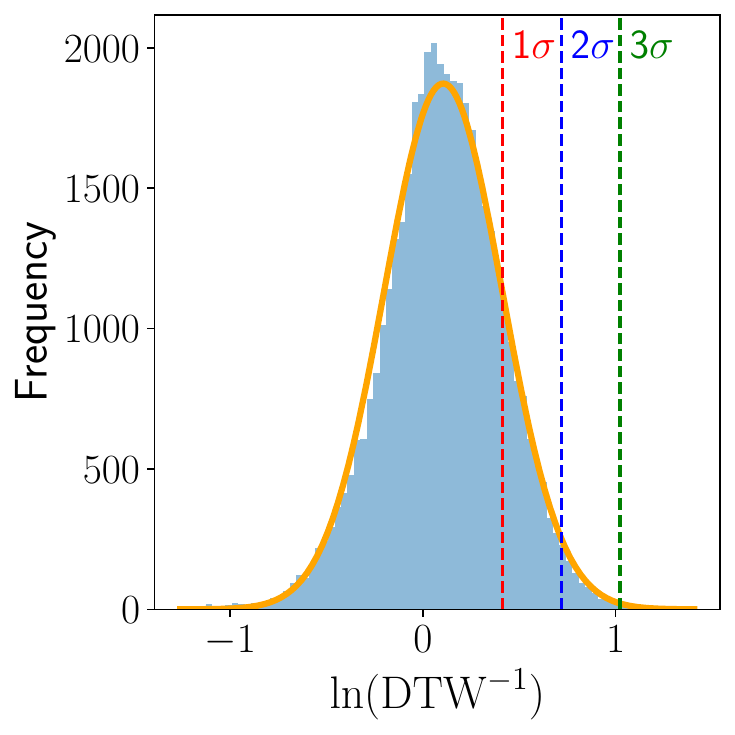}
  \centering
    \caption{
    Distribution of the natural logarithm of $\mathrm{DTW}^{-1}$ values, $\ln (\mathrm{DTW}^{-1})$. 
    The orange line shows the normal distribution identified as the best fit for the data. The dashed vertical lines denoted with $1\sigma$ (red), $2\sigma$ (blue), and $3\sigma$ (green) represent the intervals of the distribution according to the empirical rule.
    }
       \label{fig: distribution_dtw}
\end{figure}

As the DTW methodology does not need to be applied over time series of equal length, we now consider all light curves without any rebinning. 
We obtain 294 ordered rankings, one for each pulsar; we have (294 $\times$ 293)/2 = 43071 DTW values. 
The log-normal distribution is identified as the best fit for both $\mathrm{DTW}$ and $\mathrm{DTW}^{-1}$ values (we take the inverse values to facilitate looking at the most similar light curve pairs in the far end tail of the distribution).
Thus, considering the distribution of $\ln (\mathrm{DTW}^{-1})$ values, as shown in Fig. \ref{fig: distribution_dtw}, allows us to define confidence intervals in the usual way, as a Gaussian is the best fit for this distribution. 
We use the mean $\mu$ and the standard deviation ($\sigma$) of the distribution of $\ln (\mathrm{DTW}^{-1})$, equal to 0.105 and 0.306, respectively, to obtain a value of $3\sigma$ deviation equal to 2.78. That is, a value of $\mathrm{DTW}^{-1}=2.78$ represents two pairs of light curves whose similarity measured by DTW is 3$\sigma$ beyond the mean.
Beyond this value, 70 pairs of light curve comparisons are found, involving 63 pulsars (13 of them are MSPs).

\begin{figure*}
  \includegraphics[width=0.9\textwidth]{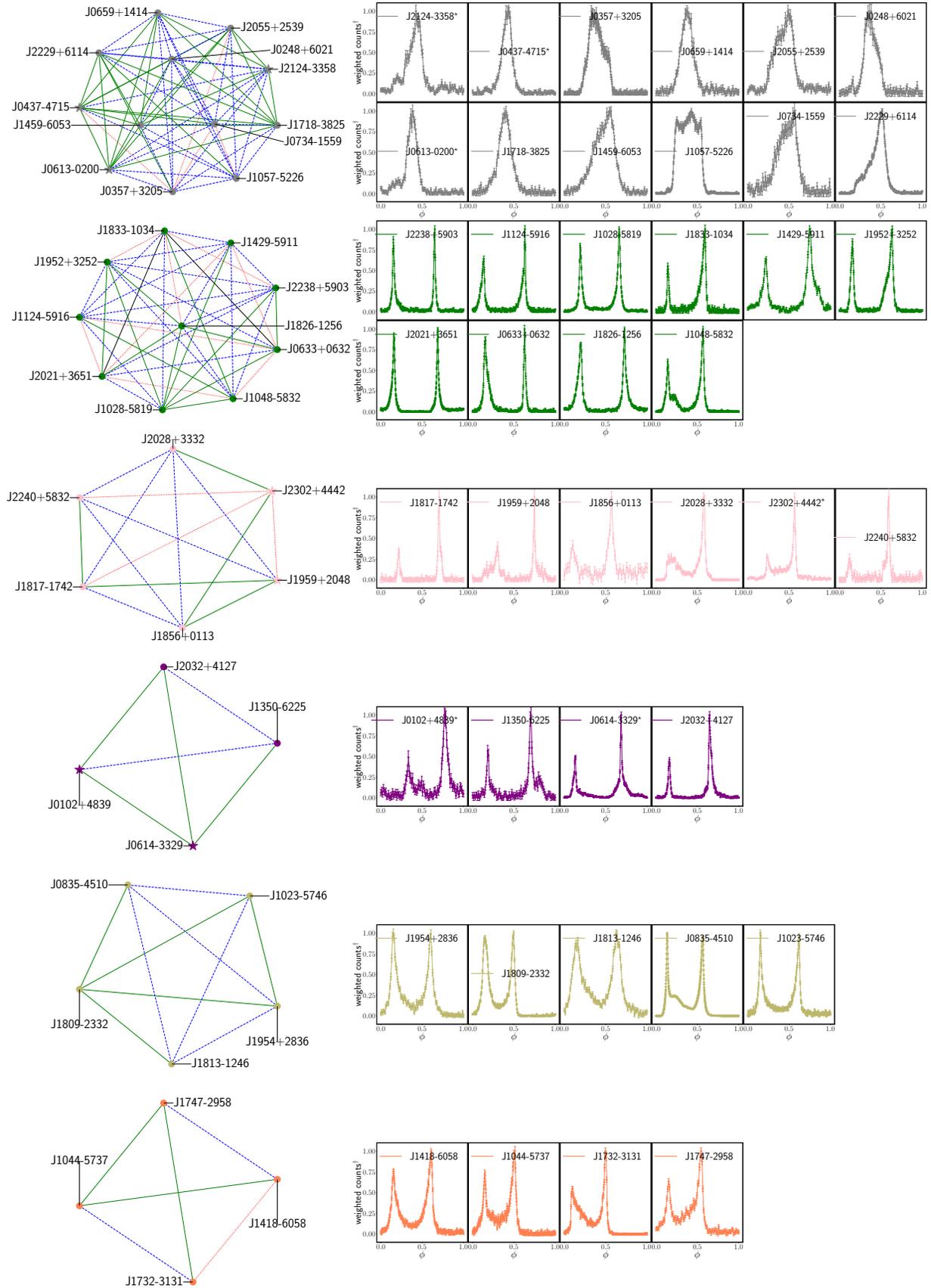}
  \centering
\caption{Clusters with 4 or more pulsars and all members related to another by at least one similarity at a 3$\sigma$ significance level. 
Each node represents a pulsar, and each colored edge connecting the nodes the similarity significance of the corresponding light curves (black for $<1\sigma$, red for cases between 1 and $2\sigma$, blue for cases between 2 and $3\sigma$, and green for $>3\sigma$).
The light curves associated with each cluster are shown next.
MSPs are denoted with a star-shaped node and on the pulsar name. }
       \label{fig: clusters_3sigma}
\end{figure*}

We can use the DTW distribution to partition the light curve sample into clusters. 
If we request clusters to contain at least 4 pulsars, and that members are related by at least one similarity at the 3$\sigma$ level, we find six clusters. 
We can depict these connections in a graph.
Figure \ref{fig: clusters_3sigma} shows these six clusters, where 41 pulsars are displayed as the graph nodes. 
The properties of these pulsars are shown in Table \ref{tab: pulsar_parameters}.
The different similarity levels according to the distribution of $\mathrm{DTW}$ values are shown by the color-coded edges, in the corresponding complete cluster graphs (where all nodes are connected). 
Figure \ref{fig: clusters_3sigma} also shows the light curves of the members of each cluster.
The complete graphs show that the vast majority of the light curves within a given cluster also belong in the tail of the distribution, with similarities beyond $>1$ and $>2 \sigma$.

\begin{table*}
\centering
\scriptsize
\caption{The columns show the pulsar timing, derived properties using the rotating dipole model, and gamma-ray properties of the clusters shown in Fig. \ref{fig: clusters_3sigma}.
The color coding is the same as that figure.
MSPs are denoted with a star on the pulsar name, and their corresponding $\dot P$ has an additional factor $10^{-6}$ multiplying the column label.
J1856$+$0113 has TS$<$25, so no spectral fit results are provided. 
See the 3PC for more details. 
}
\begin{tabular}{l | ccccc | cc }
\hline
\textbf{ PSRJ}  & $P$ & $\dot P$  & $\dot E_{sd}$  & $B_s$ & $B_{lc}$  & $F_{E>100 {\rm MeV}}$  & $E_p$ \\
                &  [s] &  [s/s]   &  [erg/s]       &  [G]  & [G]       &  [erg/cm$^2$/s]        & [GeV] \\
                &      &   [$\times 10^{-14}$]       & [$\times 10^{35}]$& [$\times 10^{10}$] & [$\times 10^{4}$] & [$\times 10^{-11}$]\\
\hline

\textcolor{Gray}{J0248$+$6021} & 0.217 & 5.52 &   2.13 & 350.35 &  0.32 &  2.94 &           0.774 \\
\textcolor{Gray}{J0357$+$3205} & 0.444 & 1.31 &   0.06 & 244.02 &  0.03 &  6.01 &           0.869 \\
\textcolor{Gray}{J0437$-$4715$^{\star}$} & 0.005 & 5.72 &   0.12 &   0.06 &  2.80 &  1.76 & 0.700 \\
\textcolor{Gray}{J0613$-$0200$^{\star}$} & 0.003 & 0.96 &   0.13 &   0.02 &  5.57 &  3.82 & 1.156 \\
\textcolor{Gray}{J0659$+$1414} & 0.384 & 5.50 &   0.38 & 465.38 &  0.08 &  2.65 &           0.156 \\
\textcolor{Gray}{J0734$-$1559} & 0.155 & 1.25 &   1.32 & 140.95 &  0.35 &  4.57 &           0.623 \\
\textcolor{Gray}{J1057$-$5226} & 0.197 & 0.58 &   0.30 & 108.54 &  0.13 & 29.59 &           1.213 \\
\textcolor{Gray}{J1459$-$6053} & 0.103 & 2.53 &   9.08 & 163.32 &  1.37 & 12.27 &           0.372 \\
\textcolor{Gray}{J1718$-$3825} & 0.074 & 1.32 &  12.49 & 100.35 &  2.22 & 10.37 &           0.564 \\
\textcolor{Gray}{J2055$+$2539} & 0.319 & 0.41 &   0.05 & 115.85 &  0.03 &  5.32 &           1.215 \\
\textcolor{Gray}{J2124$-$3358$^{\star}$} & 0.004 & 2.06 &   0.07 &   0.03 &  2.48 &  3.88 & 1.949 \\
\textcolor{Gray}{J2229$+$6114} & 0.051 & 7.53 & 215.58 & 199.49 & 13.33 & 24.01 &           0.740 \\

\hline

\textcolor{green}{J0633$+$0632} & 0.297 &   7.96 &   1.19 &  492.24 &  0.17 &  9.56 &1.427 \\
\textcolor{green}{J1028$-$5819} & 0.091 &   1.42 &   7.34 &  115.28 &  1.39 & 24.51 &0.888 \\
\textcolor{green}{J1048$-$5832} & 0.124 &   9.55 &  19.92 &  347.86 &  1.69 & 18.52 &0.968 \\
\textcolor{green}{J1124$-$5916} & 0.136 &  75.15 & 119.15 & 1021.18 &  3.78 &  6.11 &0.362 \\
\textcolor{green}{J1429$-$5911} & 0.116 &   2.39 &   6.06 &  168.30 &  1.00 & 11.28 &0.540 \\
\textcolor{green}{J1826$-$1256} & 0.110 &  12.12 &  35.72 &  369.86 &  2.54 & 41.35 & 0.870 \\
\textcolor{green}{J1833$-$1034} & 0.062 &  20.20 & 336.24 &  357.81 & 13.89 &  8.91 &0.305 \\
\textcolor{green}{J1952$+$3252} & 0.040 &   0.58 &  37.23 &   48.56 &  7.24 & 14.77 &0.993 \\
\textcolor{green}{J2021$+$3651} & 0.104 &   9.48 &  33.53 &  317.40 &  2.62 & 49.21 &1.064 \\
\textcolor{green}{J2238$+$5903} & 0.163 &   9.69 &   8.87 &  401.71 &  0.86 &  6.64 &0.673 \\

\hline

\textcolor{Lavender}{J1817$-$1742} & 0.150 &  2.06 & 2.42 & 177.62  &  0.49 & 2.85            &  1.123 \\
\textcolor{Lavender}{J1856$+$0113} & 0.267 & 20.59 & 4.25 & 750.91  &  0.36 &  -              &     - \\
\textcolor{Lavender}{J1959$+$2048$^{\star}$} & 0.002 &  1.68 & 1.60 &   0.02  & 36.73 & 1.57  & 1.038 \\
\textcolor{Lavender}{J2028$+$3332} & 0.177 &  0.49 & 0.35 &  93.74  &  0.16 & 5.71            & 1.511 \\
\textcolor{Lavender}{J2240$+$5832} & 0.140 &  1.53 & 2.20 & 147.82  &  0.50 & 0.98            & 1.138 \\
\textcolor{Lavender}{J2302$+$4442$^{\star}$} & 0.005 &  1.33 & 0.04 &   0.03  &  1.75 & 3.90  & 2.355 \\
\hline

\textcolor{Plum}{J0102$+$4839$^{\star}$} & 0.003 & 1.17 & 0.17 &   0.02 & 6.68 &  1.50  & 1.700 \\
\textcolor{Plum}{J0614$-$3329$^{\star}$} & 0.003 & 1.78 & 0.22 &   0.02 & 7.06 & 11.46  & 2.835 \\
\textcolor{Plum}{J1350$-$6225} & 0.138 & 0.89 & 1.33 & 112.10 & 0.39 &  3.60            & 1.951 \\
\textcolor{Plum}{J2032$+$4127} & 0.143 & 1.16 & 1.56 & 130.56 & 0.41 & 14.24            & 2.606   \\

\hline

\textcolor{olive}{J0835$-$4510} & 0.089 & 12.23 &  67.64 &  334.51 & 4.31 & 929.32 &  1.267 \\
\textcolor{olive}{J1023$-$5746} & 0.111 & 37.99 & 108.21 &  658.53 & 4.37 &  14.60 &  0.621 \\
\textcolor{olive}{J1809$-$2332} & 0.147 &  3.44 &   4.29 &  227.33 & 0.66 &  42.48 &  1.285 \\
\textcolor{olive}{J1813$-$1246} & 0.048 &  1.76 &  62.39 &   92.96 & 7.70 &  24.78 &  0.483 \\
\textcolor{olive}{J1954$+$2836} & 0.093 &  2.12 &  10.48 &  141.71 & 1.64 &  10.70 &  1.172 \\
\hline

\textcolor{orange}{J1044$-$5737} & 0.139 &  5.46 &  8.02 & 278.71  & 0.95 & 11.40 & 0.650 \\
\textcolor{orange}{J1418$-$6058} & 0.111 & 17.10 & 49.92 & 440.00  & 3.00 & 29.70 & 1.001 \\
\textcolor{orange}{J1732$-$3131} & 0.197 &  2.80 &  1.46 & 237.56  & 0.29 & 17.93 & 2.089 \\
\textcolor{orange}{J1747$-$2958} & 0.099 &  6.11 & 24.98 & 248.61  & 2.37 & 15.91 & 0.526 \\

\hline
\end{tabular}
\label{tab: pulsar_parameters}
\end{table*}

\section{Conclusions}
\label{conclu}

This Letter provides a quantitative assessment of gamma-ray light curve similarity that goes beyond estimating global properties, such as the number of peaks, the peak width, or the peak separation. 
Studying via DTW the similarity of all light curves (more than 43000 pairs, each with their corresponding rotations arising by considering that phase zero is arbitrary), a well-behaved distribution is found, from which similarity can be assessed.
Referring to that distribution of DTW values, we could define intervals of confidence and state for the first time if a given pair of pulsars have light curve similarity above or below a given threshold. 
This methodology is devoid of the caveats affecting the point-by-point comparison provided by the Euclidean distance between two-time series.
The latter can only be applied to light curves described with equal resolution, and rebinning not only loses details of the morphology in those cases that are better sampled but also requires that the sample is cut unless light curves with the worst resolution drive the morphological comparison.
Even if we are to rebin and cut the sample accordingly, light curves that, for instance, have two very similar peaks with similar heights and widths but whose peak separation is slightly different would never be recalled as being similar by the Euclidean distance estimator.
DTW, instead, would clearly mark those as such since it looks at the light curves globally.
Using this methodology, we were able to cluster the light curve sample in quantitative terms.

The variety of the physical and spectral properties of the light curves that are morphologically similar is large.
Having the whole distribution of the DTW, we have, in fact, explored whether the similarity of any two light curves, as measured with the DTW technique, is correlated with nearness in any other quantity of interest. 
In particular, we tested for correlations between the $\mathrm{DTW}$ values for any two pulsars of the sample and the Euclidean distance under each one of the pulsar properties considered above.
For instance, we considered the $(\mathrm{DTW})^{-1}$ measure between two given pulsars versus the properly normalized difference of their periods or their magnetic field at the light cylinder. 
No correlation arises for any of the physical or spectral parameters.
Similar light curves do not imply that pulsars are also similar in terms of their timing properties and can, in fact, mix MSPs with young ones. 
Even for pulsars of the same kind,  similar morphologies do not imply having a similar period derivative, magnetic field, or spin-down energy; they all can vary significantly.
This suggests that gamma-ray emission is radiated by the same mechanism in all pulsars and that geometry dominates the light curve morphology.

\section*{Acknowledgements}
CRG is funded by the Ph.D FPI fellowship PRE2019-090828 and acknowledges the graduate program of the Universitat Aut\`onoma of Barcelona. This work was also supported by the Spanish program Unidad de Excelencia María de Maeztu CEX2020-001058-M and by MCIU with funding from the European Union NextGeneration EU (PRTR-C17.I1), as well as
by PID2021-124581OB-I00 funded by MCIN/AEI /10.13039/501100011033. We acknowledge D. Vigan\`o and D. I\~niguez, and the participants of the
International Pulsar Conference (Guiyang, December 2024) for discussions.


\bibliographystyle{aasjournal}
\bibliography{biblio}

\clearpage

\section*{Appendix}
\subsection*{A DTW example}
\label{appendix1}

\begin{figure}
\centering
  \includegraphics[width=0.25\textwidth]{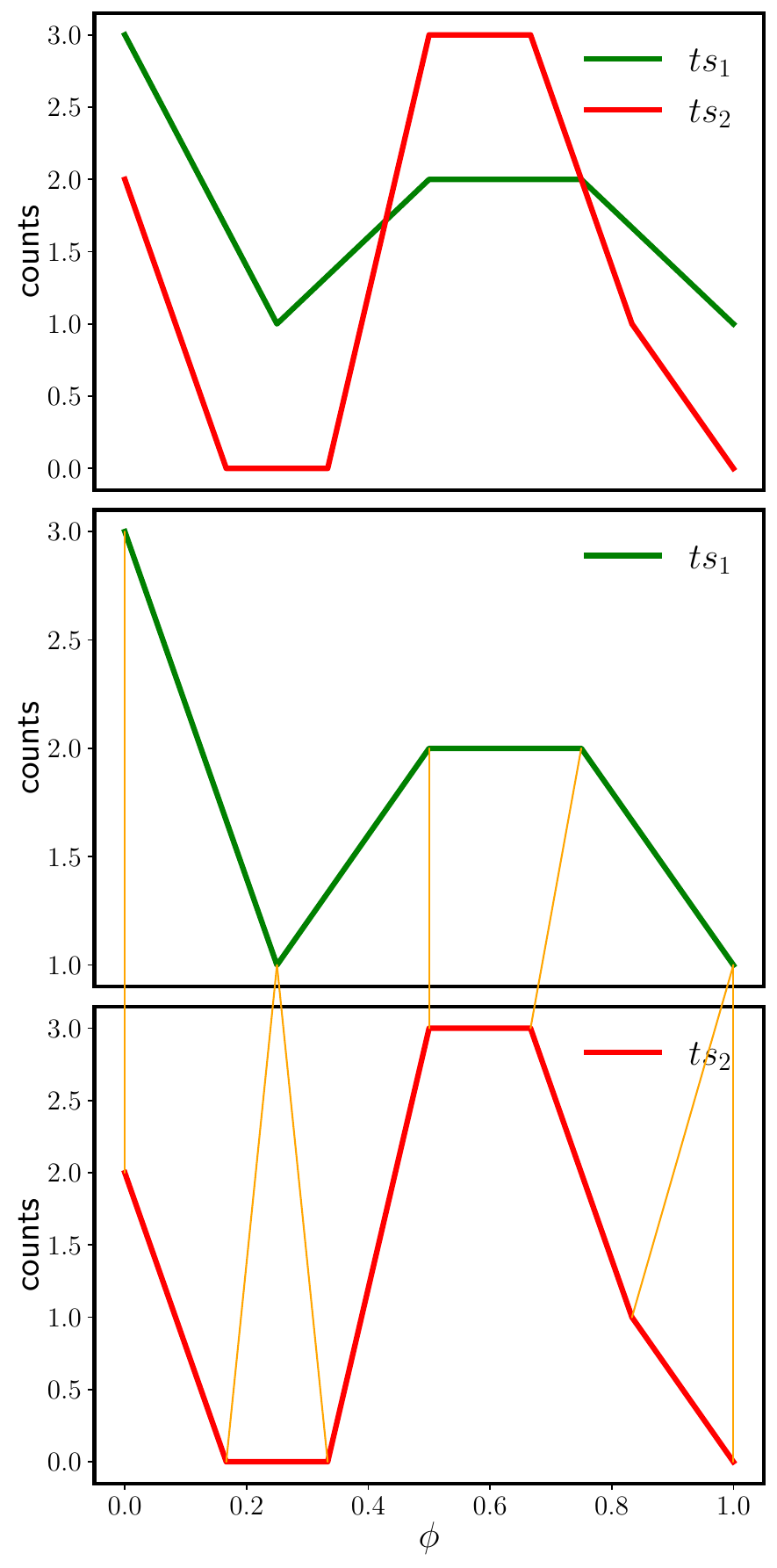}
  \caption{Top panel: 
Time series $ts_1$ (green) and $ts_2$ (yellow) used in the DTW example.
The y-axis represents an arbitrary number of counts according to Eq. (\ref{eq: toy_example}).
The x-axis represents the phases for each value, from 0 to 1, defined by the number of values in each time series.
Bottom panel:
Application of DTW. 
The orange lines represent $\pi_{o}$, the connection between the pairs of the time series (i.e., the chosen ($i, j$) pairs from Eq. (\ref{eq: MatrixDcum})).
}
  \label{fig: appendix1_tseries_DTW}
\end{figure}

Consider as an example the following two-time series, see Fig. \ref{fig: appendix1_tseries_DTW}, one with 5 and another with 7 instances
\begin{equation}
  \begin{aligned}
     ts_1&=&3, 1, 2, 2, 1\\
     ts_2&=&2, 0, 0, 3, 3, 1, 0.
    \label{eq: toy_example}
\end{aligned}  
\end{equation}
We calculate the Euclidean distance matrix ($E_{d}$), see eg., \cite{EuclideanDistance_matrix}), as
\begin{equation}
E_{d}=E(d(i,j)^{2})=\begin{bmatrix}
        1 & 9 & 9 & 0 & 0 & 4 & 9 \\
        1 & 1 & 1 & 4 & 4 & 0 & 1 \\
        0 & 4 & 4 & 1 & 1 & 1 & 4 \\
        0 & 4 & 4 & 1 & 1 & 1 & 4 \\
        1 & 1 & 1 & 4 & 4 & 0 & 1 \\
\end{bmatrix}
~.\label{eq: MatrixED}
\end{equation}
Here, each value is the square of the difference between the elements of the series located in ($i, j$), that is, $d = (ts_{1}(i)-ts_{2}(j))^{2}$, where $i$ is an index running on $ts_1$, from 1 to 5, and $j$ runs on $ts_2$, from 1 to 7.
In the next step, we compute the cumulative distance matrix ($D$) through the next recurrence relation, which is the basis of the dynamic programming:
\begin{eqnarray}
D(i,j)=E_d(i,j)+ && \nonumber \\ && \hspace{-2cm}
\min\big(D(i-1,j),D(i,j-1),D(i-1,j-1)\big)
    .\label{eq: D_cumulative}
\end{eqnarray}
In our example, the application of Eq. (\ref{eq: D_cumulative}) gives
\begin{equation}
D= D(i,j) = \begin{bmatrix}
    \textcolor{orange}{1} & 10 & 19 & 19 & 19 & 23 & 32 \\
    2 & \textcolor{orange}{2}  &  \textcolor{orange}{3} &  7 & 11 & 11 & 12 \\
    2 &  6 &  6 &  \textcolor{orange}{4} &  5 &  6 & 10 \\
    2 &  6 & 10 &  5 &  \textcolor{orange}{5} &  6 & 10 \\
    3 &  3 &  4 &  8 &  9 &  \textcolor{orange}{5} &  \textcolor{orange}{6} 
    \label{eq: MatrixDcum}
\end{bmatrix}
\end{equation}
For instance, the value seen in row 2 and column 2 in Eq. (\ref{eq: MatrixDcum}) is obtained as $D(2,2)=E_d(2,2)+ \min\big(D(1,2), D(2,1), D(1,1)\big)=1+\min(2,10,1)=2$.

The aim will be to find the optimal path (the minimal cost), referred to as $\pi_{o}$, that relates the two series.
This can be found by tracing backward (also called backtrack) in $D$, that is, taking the previous cells with the lowest cumulative values from the initial cell.
The optimization will be subject to the following conditions.
First, it needs to connect the beginning and the end of each of the time series, i.e., it will be a path, $\pi$, that starts at cell (5, 7) and ends at cell (1, 1). 
Also, we need to ensure that the path $\pi$ respects the sequential order of the time series, so the path should move through the cells either to the left or upward, never to the right or downward. For instance, from cell (5, 7), the next lowest value is 5, which occupies the cell (5, 6).
Finally, to avoid skipping any element of the time series, each step in $\pi$ is constrained to reach neighboring cells only. For instance, the next cell to reach from (5, 6) can only be the cells (4, 5), (4, 6), or (5, 5), whose values, respectively, are 5, 6, and 9, thus $\pi$ goes through (4, 5) which is the cell with the minimum cost.
When there is more than one possible cell with the same value, the one in the diagonal is prioritized as the most natural one-to-one alignment.
In this case, the colored values seen in Eq. (\ref{eq: MatrixDcum}) represent $\pi_{o}$, which connects the elements (1, 1), (2, 2), (2, 3), (3, 4), (4, 5), (5, 6) and, (5, 7).
Its final cost, i.e., the DTW value, is the square root of the sum of each of the above-selected elements, according to Eq. (\ref{eq: MatrixED}) and results in 2.45. 
Visually, $\pi_o$ is shown in the bottom panel of Fig. \ref{fig: appendix1_tseries_DTW}, where each orange line represents the connection of each of the elements extracted from Eq. (\ref{eq: MatrixDcum}).

\newpage

\subsection*{The Euclidean path}
\label{appendix2}

Consider the following time series, having the same length
\begin{eqnarray}
\begin{aligned}
     ts_3&=&3, 1, 2, 2, 1\\
     ts_4&=&2, 0, 0, 3, 3
    \label{eq: toy_example2}
\end{aligned}
\end{eqnarray}

Calculating $E_{d}$ and $D$, as shown before we get
\begin{equation}
E_{d}= \begin{bmatrix}
    1 & 9 & 9 & 0 & 0 \\
    1 & 1 & 1 & 4 & 4 \\
    0 & 4 & 4 & 1 & 1 \\
    0 & 4 & 4 & 1 & 1 \\
    1 & 1 & 1 & 4 & 4 \\
\end{bmatrix}
~,\label{eq: MatrixED2}
\end{equation}
and
\begin{equation}
D = \begin{bmatrix}
    \colorbox{blue}{\textcolor{orange}{1}} & 10 & 19 & 19 & 19 \\
    2 &  \colorbox{blue}{\textcolor{orange}{2}} &  \textcolor{orange}{3} &  7 & 11 \\
    2 &  6 &  \colorbox{blue}{6} &  \textcolor{orange}{4} &  5 \\
    2 &  6 & 10 &  \colorbox{blue}{\textcolor{orange}{5}} &  5 \\
    3 &  3 &  4 &  8 &  \colorbox{blue}{\textcolor{orange}{9}} 
\end{bmatrix}
~.\label{eq: MatrixDcum2}
\end{equation}

\begin{figure} 
\centering
  \includegraphics[width=0.49\textwidth]{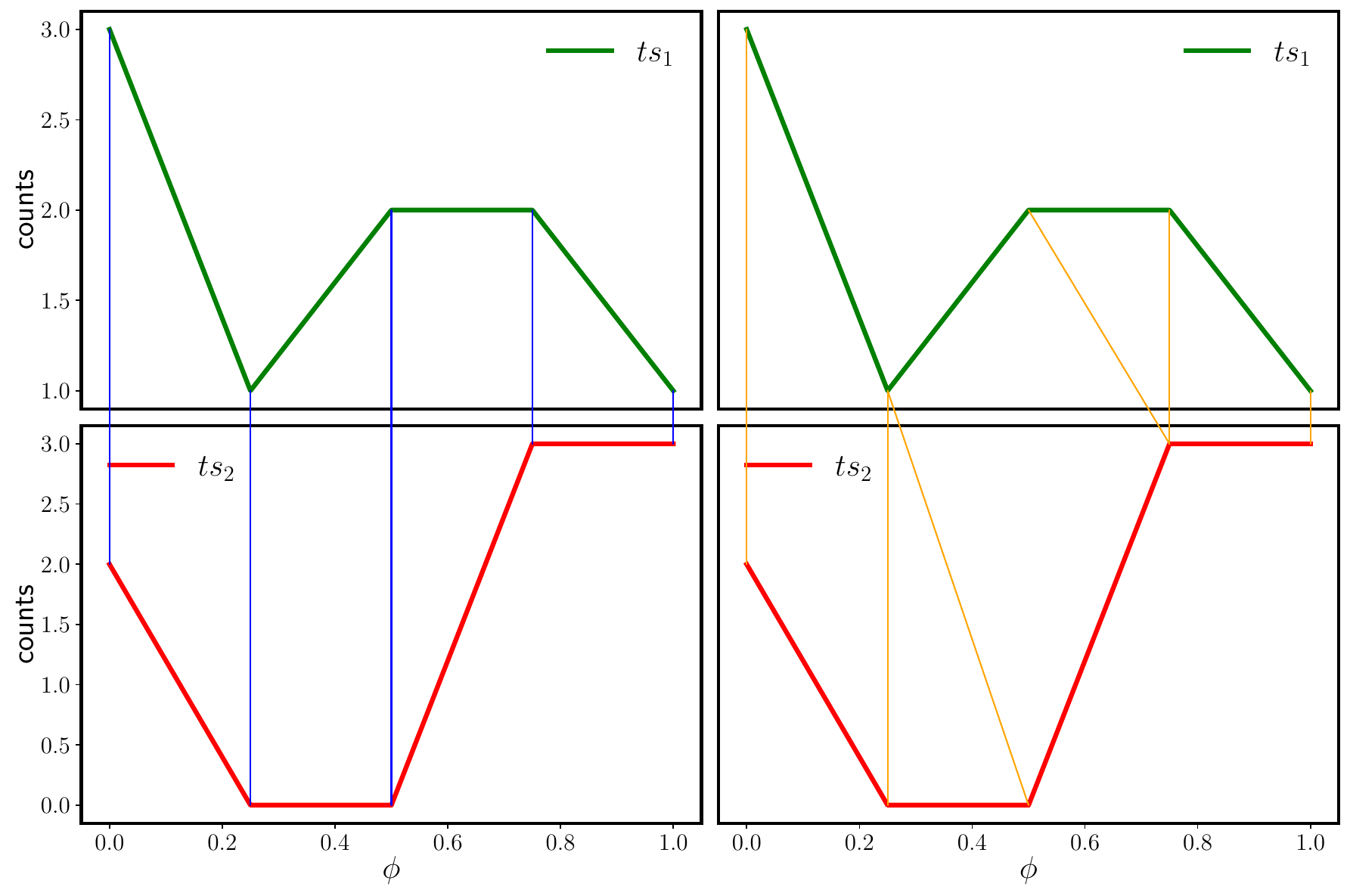}
  \caption{Time series $ts_3$ (green) and $ts_4$ (yellow) seen in Eq. (\ref{eq: toy_example2}) connected via the Euclidean alignment (left) or DTW (right).
}
  \label{fig: appendix2_DTW_ED}
\end{figure}

The main diagonal in Eq. (\ref{eq: MatrixDcum2}) (noted in blue) represents a $\pi$ that has been explored in computing Eq. (\ref{eq: DTW_optimized}), arising by the connection of all $(i,j)$ with $i=j$; a point-by-point comparison of the time series.
According to Eq. (\ref{eq: MatrixED2}), its cost would be 3.31.
However, when applying Eq. (\ref{eq: DTW_optimized}), we see that another $\pi$, that would involve a different set of pairs $(i,j)$, and has a lower cost, being $\pi_{o}$ in Eq. (\ref{eq: MatrixDcum2}) (noted in orange).
According to Eq. (\ref{eq: MatrixED2}), the cost of the blue path would be 3.00.
In Fig. \ref{fig: appendix2_DTW_ED}, the alignment through the set ($i,j$) following $\pi$ and $\pi_{o}$, corresponding to both methods, is shown with the same color code used in Eq. (\ref{eq: MatrixDcum2}).
Therefore, if the time series have the same length, the direct alignment between the two series (also referred to as the Euclidean alignment or as just the Euclidean distance (ED) of the time series) will be computed and compared in the search for the minimal value.
This may or not be the $\pi_o$, and in this example, it is not.

\subsection*{Implementation and computational cost}
\label{appendix3}

The work has been done in Python v3.11.2, using the dtaidistance v2.3.10 package (\cite{DTW_package}) for the DTW application. 
The number of DTW values for each light curve comparison, DTW($lc_1$, $lc_2$), depends on the $N$ of each light curve.
Due to the applied rotation being $N_1$, and $N_2$, the size corresponding to $lc_1$ and $lc_2$, respectively, to this comparison is $N_{1}\times N_{2}$. 
Considering the 43071 possible pairs, the $N$ of each light curve, the number of computations is higher than $3\times 10^8$.
The implementation has been carried out on the Hidra High-Performance Computing (HPC) Cluster hosted at the Institute of Space Sciences (ICE, CSIC).
As an example, in a comparison involving light curves of 25 bins and 50 bins, the time usage required for a single (bin-to-bin) DTW computation is $\sim$ 0.0014~s, while the total time considering the shift is 1.407 s. In this case, the memory usage is negligible, 0.78 MB.
For a comparison taking light curves of 100 bins and 200 bins, the time usage for a single DTW is $\sim$ 0.0190 s, with a total of 371.186~s if looking at the full rotation set. 
The memory usage is around 3.75 MB. The computations were distributed using multiprocessing along 16 cores for a more efficient process.

\end{document}